# Optoelectronic characteristics and application of black phosphorus and its analogs

Ying-Ying Li, Bo Gao*, Ying Han, Bing-Kun Chen, Jia-Yu Huo

College of Communication Engineering, Jilin University, Changchun 130012, PR China

Corresponding authors. E-mail: *gaobo0312@jlu.edu.cn

## Abstract

The tunable bandgap from 0.3 eV to 2 eV of black phosphorus (BP) makes it to fill the gap in graphene. When studying the properties of BP more comprehensive, scientists have discovered that many two-dimensional materials, such as tellurene, antimonene, bismuthene, indium selenide and tin sulfide, have similar structures and properties to black phosphorus thus called black phosphorus analogs. In this review, we briefly introduce preparation methods of black phosphorus and its analogs, with emphasis on the method of mechanical exfoliation (ME), liquid phase exfoliation (LPE) and chemical vapor deposition (CVD). And their characterization and properties according to their classification of single-element materials and multi-element materials are described. We focus on the performance of passively mode-locked fiber lasers using BP and its analogs as saturable absorbers (SA) and demonstrated this part through classification of working wavelength. Finally, we introduce the application of BP and its analogs, and discuss their future research prospects.

## 1. Introduction

In recent years, two-dimensional layered materials (2DLMs) represented by graphene have attracted wide attentions from scientists all over the world [1-6]. However, even though graphene has many excellent properties, its zero-bandgap properties largely hinder its applications such as in semiconductors as logic switches [7, 8]. The appearance of black phosphorus makes up for this deficiency [9-11]. Black phosphorus has excellent optoelectronic and nonlinear optical properties similar to graphene, as well as a tunable bandgap from 0.3 eV to 2 eV depending on the number of layers [12, 13], which allows it to have quite large modulation depth and very flexible working range.

In 1914, Bridgman reported a novel method to fabricate black phosphorus at a moderate temperature (200 °C) and a high pressure (1.2 GPa) [14]. Since then, the research on BP has not made substantial progress. Until 2014, two groups of scientists, one from the United States and the other from China, successfully exfoliated black phosphorus down to two or three atomic layers [15]. After that, the study of the preparation, properties and applications of two-dimensional (2D) black phosphorus has become a hot topic [15-21]. While the mystery about the properties of black phosphorus was gradually revealed by scientists, they found that some materials have similar molecular structure and optoelectronic properties with black phosphorus [9, 22-24]. They simply named these materials as the black phosphorus analogs (i.e. tellurene, antimonene, bismuthene, indium selenide and tin sulfide) [25, 26].

With the rapid development of the laser field, scientists have recently paid more attention to the nonlinear optical properties of black phosphorus and its analogs, which allow them to be used in passively mode-locked fiber lasers as saturable absorbers to generate shorter pulses [27-32]. The basic thought of using saturable absorbers for pulse mode-locking is that when the light pulse passes through the absorber, the loss of the edge part is greater than the loss of the central part (which is strong enough to saturate the absorber), the light pulse is thus narrowed [33, 34]. According to above reasons, the parameters of saturable absorber play an important role in the generation of ultrashort pulses. Scientists noticed that black phosphorus and its analogs have outstanding and unique optical properties such as low saturation intensity, fast carrier mobility and so on [35-38]. To be more specific, large modulation depth and fast inter-band relaxation time are helpful to obtain narrower pulse in mode-locking fiber lasers. The low non-saturable absorption loss can effectively reduce the loss in the cavity and is more conducive to the formation of the pulse in the cavity. Meanwhile, black phosphorus and its analogs generally have the advantage of being small in size, which makes them easy to make into various forms of saturable absorbers to meet diverse needs.

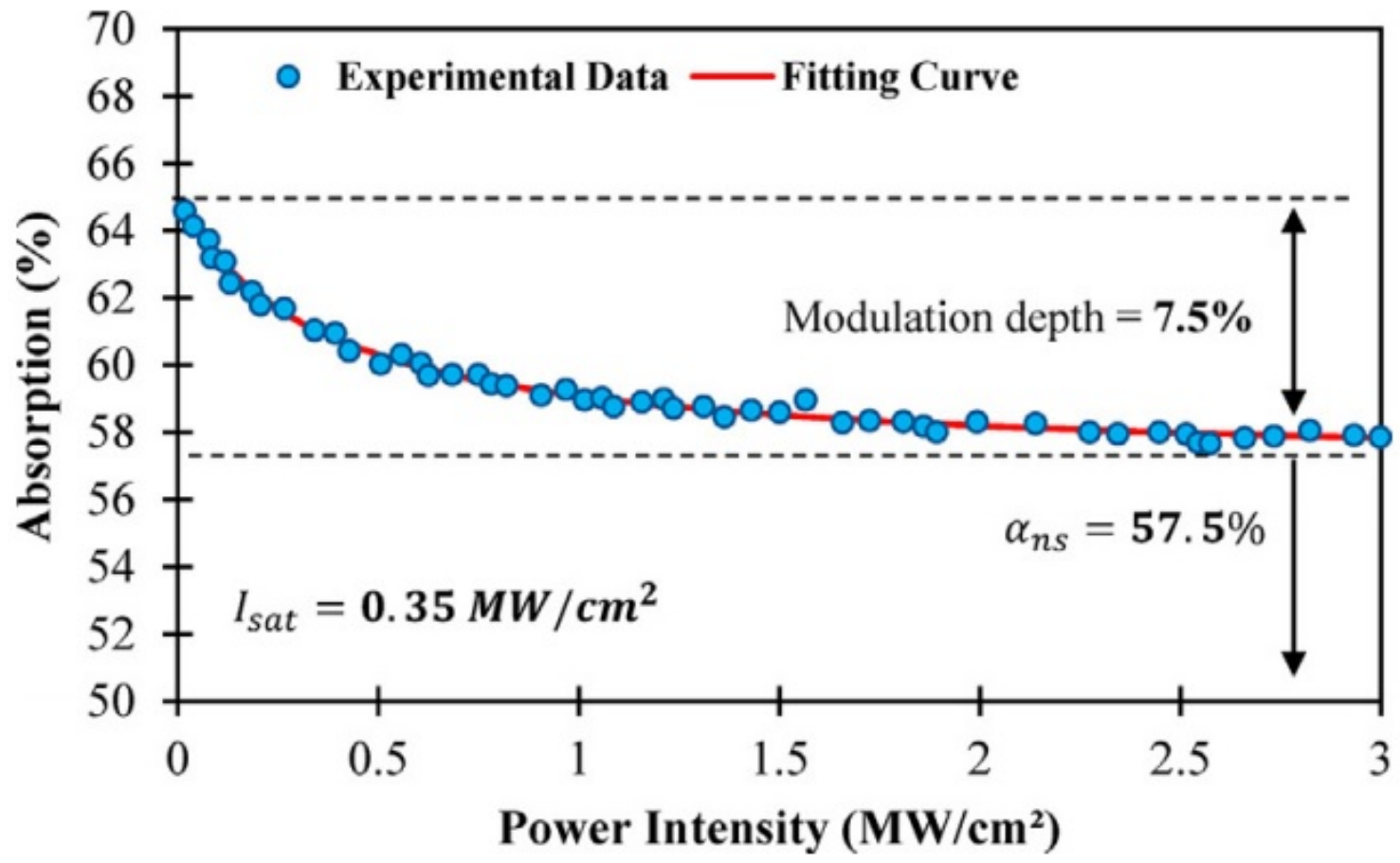

Fig. 1 Nonlinear absorption profile of BP SA. Reproduced with permission from Ref. [39].

Scientists have established models based on the electronic transition theory and two-stage band theory to better understand the nonlinear saturation absorption phenomenon of real saturable absorbers. This model is usually used to fit the data when calculating the parameters of saturable absorbers. The function of transmittance and incident light intensity is shown as follows.

$$\alpha(I) = \alpha_s/(1 + I/I_{sat}) + \alpha_{ns}$$

where α(I) is absorption coefficient, $\alpha_s$ presents modulation depth, $I_{sat}$ is the saturation intensity and $\alpha_{ns}$ is the non-saturable absorption loss. As shown in Fig. 1, Arof et al. [39] measured the nonlinear absorption curve of the saturable absorber based on BP using absorption technique. From the fitting curve, we can clearly observe that the modulation depth, saturation intensity and non-saturable absorption are 7.5%, 0.35 MW/cm$^2$, 57.5%, respectively. Strong aspiration of better performance has always been at the central issue of experts [40-46]. Li et al. [47] obtained few-layer black phosphorus saturable absorbers with non-saturable absorption of 1% and saturation intensity of 221 MW/cm$^2$. Sb, a mono-element 2D material, has been reported to generate ultrashort pulse in Tm-doped fiber lasers as SA with the modulation depth of 48% [48]. Wang et al. [36] reported a bismuthine-based SA with the saturation intensity of 113 MW/cm$^2$.

This paper is organized as follows: in section 1, we will give a brief introduction and background on black phosphorus and its analogs. In section 2, we will introduce the fabrication method of black phosphorus and its analogs. The method of mechanical exfoliation, liquid phase exfoliation and chemical vapor deposition will be elucidated in detail. In section 3, we will introduce the characterization of black phosphorus and its analogs one by one. In section 4, we will discuss the performance of black phosphorus and its analogs as saturable absorbers in passively mode-locking fiber lasers working at 1, 1.5, 2, 3 μm. In section 5, the applications of black phosphorus and its analogs will be elucidated. We described the applications of optoelectronics, electronics, biomedicine and other fields respectively. In section 6, we will give some conclusions of this paper and several outlooks about black phosphorus and its analogs.

## 2. Fabrication methods of Black phosphorus and its analogs

An increasingly number of researchers have paid attentions on the synthesis of black phosphorus and its analogs due to their excellent physical properties and wide application. At present, there are two main fabrication methods, one is called the top-down method, and the other is called the bottom-up method [49-51]. The former includes mechanical exfoliation (ME) [52-54], liquid phase exfoliation (LPE) [27, 43, 55], electrochemical exfoliation [51, 56], thermal annealing [57], plasma thinning [58, 59] and sonochemical exfoliation method [60-63]; the latter includes chemical vapor deposition (CVD), pulsed laser deposition [64-67], hydrothermal method [66], van der Waals epitaxy [68, 69] and magnetron sputtering deposition (MSD) [48, 70]. We will briefly introduce three common preparation methods: ME, LPE and CVD.

### 2.1 Mechanical exfoliation

Because the layers of two-dimensional layered materials are stacked by inter-planar van der Waals force [9], Novoselov et al. [71] found a method to fabricate few-layered crystals through bulk crystals by overcoming the van der Waals force of materials, which is called mechanical exfoliation. Mechanical exfoliation is the process by which bulk crystals are gradually exfoliated into few-layered crystals layer by layer through adhesive tape [72]. It is so convenient that people usually use this method to fabricate black phosphorus in laboratory. However, the defects such as uncontrolled quality and low yield prevent the application of this method in mass production [73, 74]. In addition, even in laboratory scale studies, people cannot effectively control the thickness, size, and shape of the production by this method [72, 75]. Ismail et al. [76] fabricated α-$ln_2Se_3$ by ME method and used it as SA of the mode-locked thulium-doped fluoride fiber laser. In order to fabricate a few-layered $ln_2Se_3$, they cut a 1 mm$^2$ sample out of the $ln_2Se_3$ crystal and then placed the sample between two clear tapes for repeated exfoliation. Fan et al. [53] considered the advantages of simple and reliable process of ME and the absence of chemical reactions and expensive laboratory equipment, choosing this method to prepare BP flakes.

### 2.2 Liquid phase exfoliation

In order to achieve the large-scale exfoliation of BP and its analogs, researchers explored another method named liquid phase exfoliation to prepare them. LPE is a popular method to synthesis saturable absorbers using two-dimensional materials, and its steps are as follows [77]. Firstly, we can destroy the van der Waals forces between the layers of bulk materials in some way such as oxidation, ion intercalation [78] and ultrasonic exfoliation [78, 79]. Then we can use liquid dispersion solvent to transfer the exfoliated material to water by centrifugation. Obviously, the choice of solvent plays a key role in the preparation of materials using this technique. Currently, the commonly used solvents are ethanol [31], NMP (N-methyl-2-pyrrolidone) [72, 79-81], DMF (dimethyl formamide), IPA (isopropanol) [78], PVA (polyvinyl alcohol) [82-84], acetone, and surfactant sodium cholate. For example, Chu et al. [79] studied a LPE strategy using NMP to fabricate BP with water stability and controllable layer and dimension. Yu et al. [78] explored a novel LPE method, with the assistance of $Li_2SiF_6$, which combines ion intercalation with ultrasonic treatment to achieve a yield of up to 75%. Zhang et al. [85] produced two-dimensional nonlayered Te using IPA as solution through LPE method. Although the yield of LPE method is relatively high, large-scale production is always a problem we are confronted with. After all, ultrasonication is still a laboratory-based process [86].

### 2.3 Chemical Vapor Deposition

Chemical vapor deposition (CVD) is a method for large-scale preparation of two-dimensional materials [73], which is often used in integrated electronic devices and transparent electrodes [86]. The preparation of BP using CVD has been extensively studied. The principle of CVD method is that after the material is crushed or even vaporized, solid depositions are generated on the substrate material to form thin films through the redox [52, 87]. Obviously, the certain substrate plays an important role in the preparation using CVD. At present, the substrate materials commonly include Cu, h-BN [64] and Si [87]. Ji et al. [87] reported the main process of CVD method and obtained the BP films with average areas >3 $\mu m^2$ and ~4 layers. The bottom-up approach allows direct synthesis of 2DLMs at the molecular level, as opposed to the top-down approach. CVD methods are commonly used to directly produce high quality 2DLMs. Because temperature, air pressure and air flow will affect the quality of the few-layered materials prepared by CVD, we can control the quality of film by changing these parameters. In general, the quality of 2DLM prepared by CVD method is higher, but the corresponding cost and complexity are also higher.

### 2.4 Other Methods

In addition to the three methods described above, other methods of preparing black phosphorus and its analogs have been explored by the scientific community [57, 88-92]. As shown in Fig. 2, we introduced several main fabrication methods.

The significant advantages of electrochemical exfoliation are simple, economic and environmentally friendly. Wu et al. [93] used the BP as a cathode for electrochemical exfoliation and obtained BP nanosheets with a lower degree of oxidation than the initial material. Through thermal annealing method, Feng et al. [94] obtained large transverse sized and defect-free BP flakes and Zhang et al. [95] fabricated surface-modified InSe nanosheets with enhanced stability and photoluminescence. Lee et al. [58] obtained BP flakes with smooth surface and a designated number of layers using plasma thinning method. This method has the advantages of low temperature, high speed and high efficiency. However, because explosive gases are involved in the process, this method is dangerous and costly [59]. Lau et al. [96] reported a method named pulsed laser deposition to fabricate α-BP flakes, which have tunable direct bandgap. Zeng et al. [97] successfully synthesized high quality multilayer antimony monocrystals on mica substrate by van der Waals epitaxy. Liu et al. [98] used a novel method named “pressure quenching” to fabricate BP flakes from red phosphorus through a controllable phase transition under conditions below 0.4 GPa and 580 ℃. Wang et al. [99] synthesized Te nanosheets from a culture of a tellurium-oxyanion respiring bacteria.

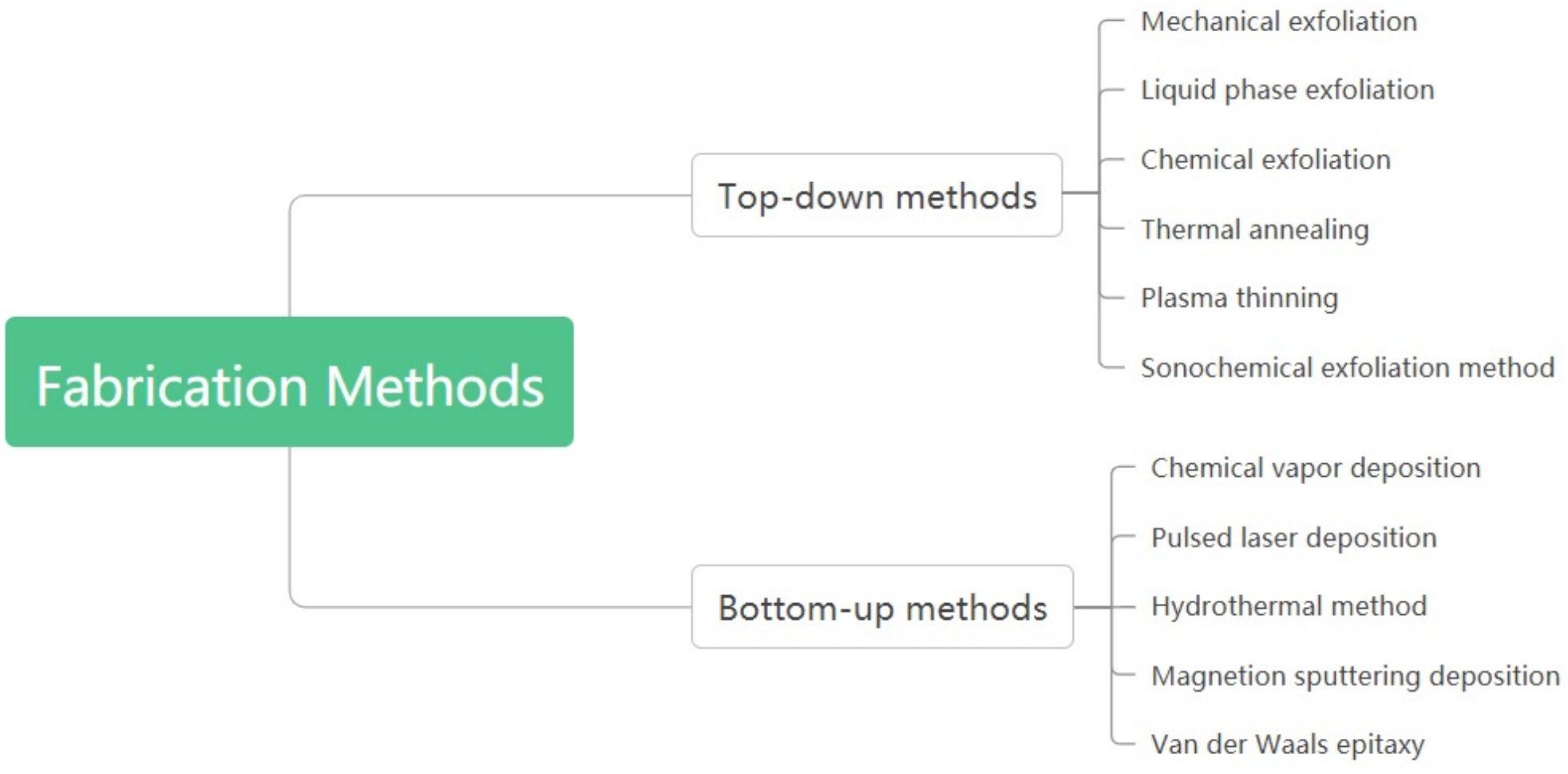

Fig. 2 The fabrication methods of BP and its analogs.

## 3. Characterization of Black Phosphorus and its Analogs

In the process from the successful preparation of a new material to the application of this material in the production, the first step is to study the properties of the material. In this section, we will introduce the characterization of black phosphorus and its analogs (including tellurium, selenium, antimony, bismuth, indium selenide, and tin sulfide). At present, the commonly used technical means to explore the properties of materials include high resolution transmission electron microscope (HRTEM), diffraction of x-rays (XRD) and Raman spectrum, etc. HRTEM image can analyze the microstructure of the sample to prove the single crystal properties of the material, XRD diagram can reflect the characteristics of elements and the distribution of atoms in the crystal, and Raman spectrum can analyze the molecular structure.

### 3.1 Black Phosphorus

Phosphorus is widely distributed on earth and is among the top 10 in the earth's crust. Phosphorus has several kinds of allotropes, such as white, red, and black phosphorus. Among them, black phosphorus is the most stable allotrope. BP has four kinds of crystal structure including orthogonal, rhombic, simple cubic and amorphous [100]. At normal temperature and pressure, BP demonstrates a unique orthogonal structure, its space group is Cmca, and the lattice constant is a =4.374 Å, b =3.3133 Å, c =10.473 Å [101]. There are eight phosphorus atoms in each cell of the black phosphorus crystal, and each atom is connected with its three adjacent phosphorus atoms through 3P hybridization orbital. The phosphorus atoms in the same layer are connected by covalent bonds, and the layers are connected by van der Waals forces [72], with an interval of 5.3 Å [81]. Meanwhile, each atom has a lone pair except for the covalent bond, which causes the air instability of BP [102]. Similar to graphene, BP is a kind of layered material and two-dimensional BP is called phosphoene [103]. However, due to different bond angles (α=96.3°, β=102.1° [104]), the phosphorus atoms in the same layer are not in a plane, but in a folded honeycomb structure, as shown in Fig. 3(a) and thus exhibit strong in-plane anisotropy between the armchair and the zigzag directions [105], as shown in Fig. 3(b). What is even more remarkable is the optical and electrical properties of BP. BP have a direct bandgap depending on thickness as shown in Fig. 3(c) [106, 107]. The bandgap of monolayer, double, triple and bulk black phosphorus are 2 eV, 1.3 eV, 0.8 eV, and 0.3 eV, respectively [105]. The bandgap of BP is the middle of graphene (0 eV) and Transition metal dichalcogenides (TMDCs) (1-2.5 eV) [108, 109]. Furthermore, Jia et al. [110] measured 15-nm-thick BP and found its field-effect mobility is up to 205 $cm^2V^{-1}s^{-1}$. Raman spectroscopy is often used to study the molecular constructure of substances. As shown in Fig. 3(d), the Raman spectroscopy of BP clearly illustrate three typical vibration modes, including 361.8 $cm^{-1}$ ($A_g^1$), 439.2 $cm^{-1}$ ($B_{2g}$) and 466.8 $cm^{-1}$ ($A_g^2$).

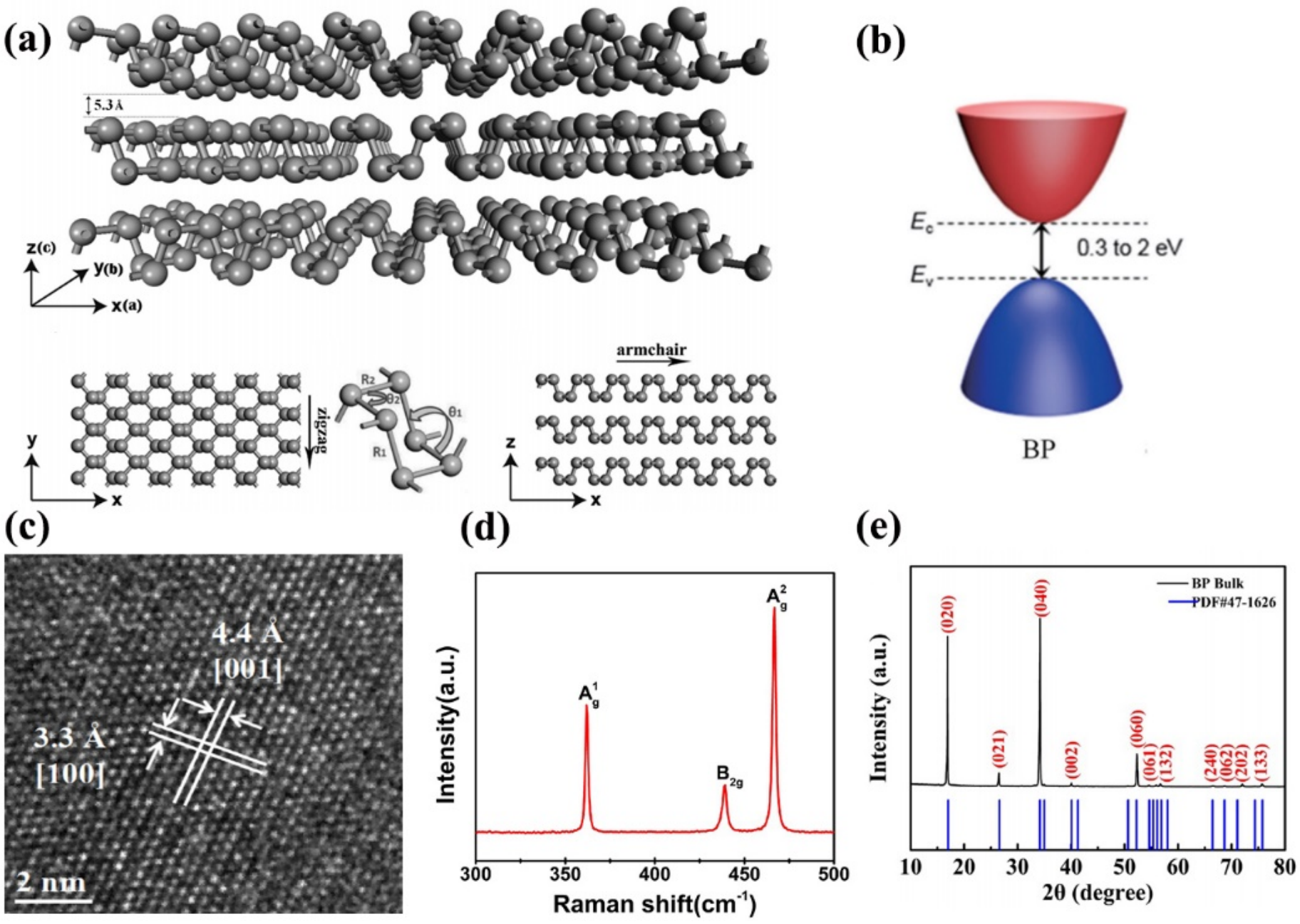

Fig. 3 (a) Folded honeycomb structure of BP. (b) Schematic diagram of bandgap of BP. (c) The HRTEM image of BP. (d) Raman spectroscopy of BP. (e) The XRD image of BP. Reproduced with permission from Ref. [72] (a), Ref. [111] (b) and Ref. [112] (c)-(e).

## 3.2 Mono-elemental analogs

### 3.2.1 Tellurene

Tellurium (Te) is one of the Group-VIA elements [113]. At present, as a new two-dimensional single-atom material, it has been used in several field such as field-effect transistors, optical detector and so on. As with other two-dimensional materials, Te has several allotropes, which includes α-Te, β-Te and γ-Te, among them, the first two being semiconductor properties and the latter exhibiting metallic properties [114]. At normal temperatures and pressures, the lattice constants of the three allotropes of Te mentioned above are shown in Table 1. Meanwhile, Te has a triangular lattice structure [100]. There are three Te atoms in each cell of Te crystal. Fig. 4(a)-(c) show the molecular structure of Te and presents us the chain-like structure of Te: each Te atom on the same chain is connected to its two adjacent Te atoms by covalent bond [100]. Therefore, as a non-layered material, Te has strong covalent bond within the chain and shows the Van Der Waals force between the chains, which is different from the inter-layer force of black phosphorus [85]. At the same time, Te has a very strong anisotropy similar to is similar to black phosphorus and other analogs, which makes it have a strong tendency to form one-dimensional nanostructures along the C axis, including Te nanorods, Te nanotubes, Te nanorires, and Te nanoribbons [115]. As shown in Fig. 4(e), the Raman spectroscopy of Tellurene shows three distinct peaks: $E_1$, $A_1$, $E_2$. Unfortunately, there are few studies on Te. In 2017, Zhang et al. [114] firstly have been focus on the environmental stability of two-dimensional Te, and they found that Te had a high carrier mobility. In 2019, Xu et al. [116] first used Z-scanning technology to explore the broadband nonlinear absorption characteristics of Te.

Table 1 the lattice constants of α-Te, β-Te and γ-Te.

| | |
|---|---|
| α-Te | a=b=4.15 Å |
| β-Te | a=4.17 Å, b=5.49 Å |
| γ-Te | a=b=3.92 Å |

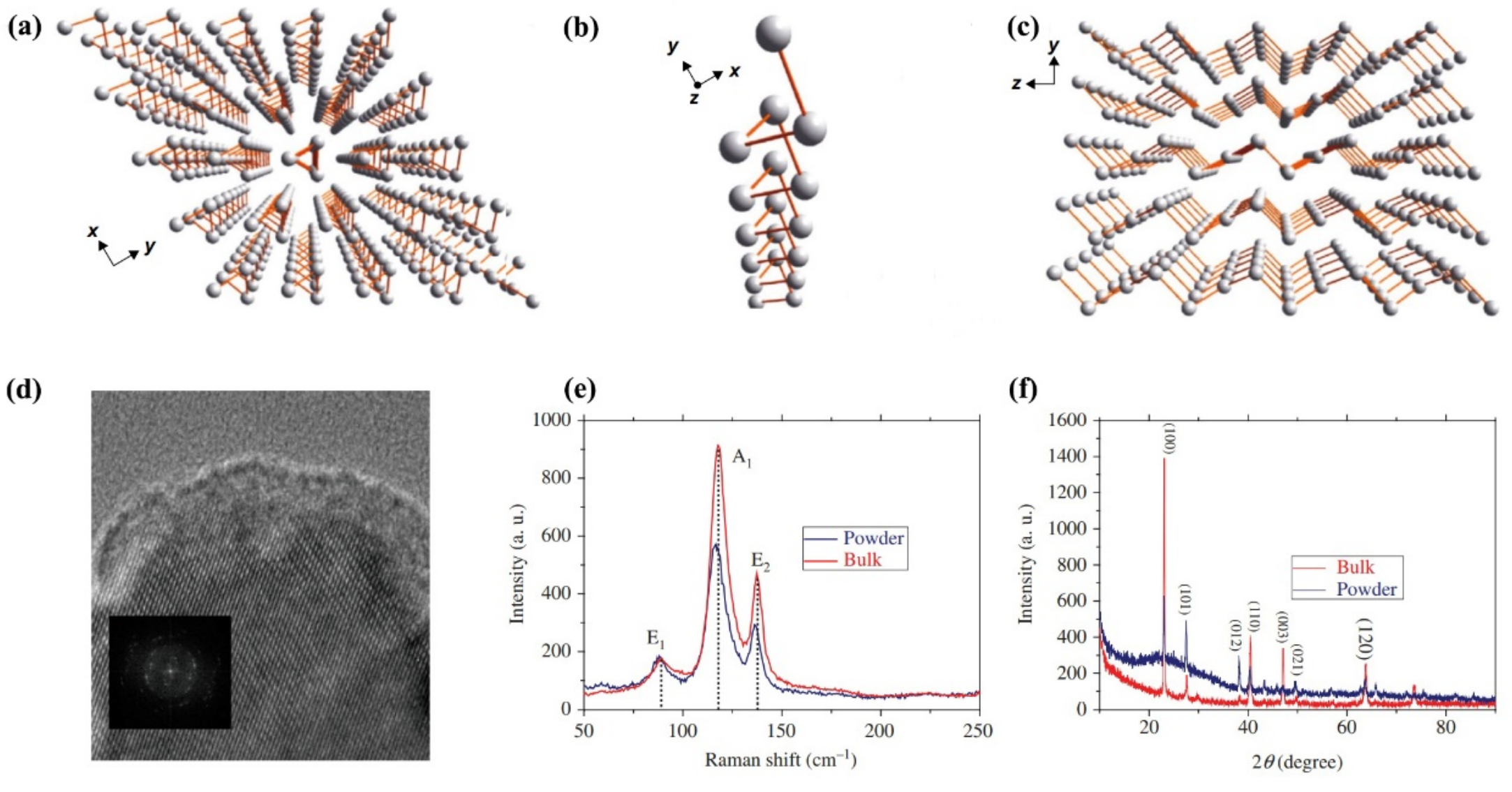

Fig. 4 (a)-(c) Crystal structure of Te viewed from the z axis, as a single-molecule chain, and viewed from the x axis. (d) The HRTEM image of Tellurene. (e) Raman spectroscopy of Tellurene. (f) The XRD image of Tellurene. Reproduced with permission from Ref. [117] (a)-(c), Ref. [99] (d) and Ref. [118] (e)-(f).

### 3.2.2 Antimonene

Antimony (Sb), like bismuth and phosphorus, belongs to the mono-element materials of Group-VA [119]. Antimony is called the cousin of phosphorus because of its similar structure and properties with phosphorus. Antimony has a dull gray metallic luster, showing a semi-metallic character [100]. According to report, there are four allotropes of antimony [120], among which β-Sb has been well studied because of its excellent stability, and its structure is shown as Fig. 5(b). Meanwhile, Fig. 5(a) shows the structure of α-Sb. Antimony has a bandgap ranging from 0 to 2.28 eV (monolayer) and can change from indirect to direct under biaxial strain [80, 121]. In addition, antimony has a buckled honeycomb structure and exhibits outstanding properties such as strong spin-orbit coupling (SOC), excellent thermal conductivity and high carrier mobility. These excellent properties of antimony have attracted scientists to further study it. Fu et al. [122] first implemented dual wavelength mode-locking using antimonene-based saturable absorbers, which was a solid step forward in the application of the generation of Terahertz waves using antimonene. Ruan et al. [48] firstly used 2D antimony as SA in ultrafast passively mode-locking fiber lasers working at 1.5 and 2 μm wavelengths.

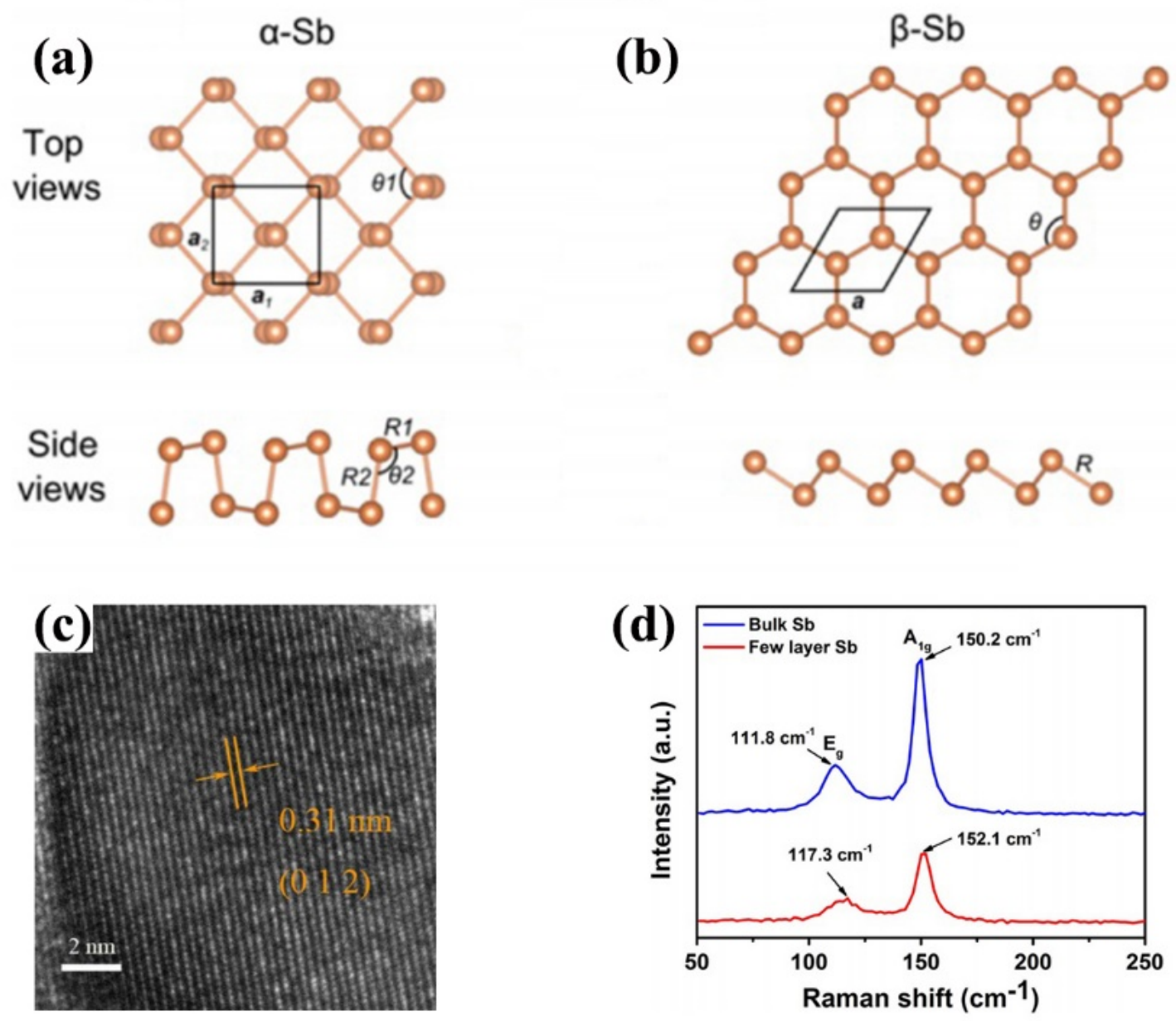

Fig. 5. (a)-(b) The structure of α-Sb and β-Sb. (c) The HRTEM image of antimonene. (d) Raman spectroscopy of antimonene. Reproduced with permission from Ref. [120] (a)-(c), Ref. [121] (c) and Ref. [121] (d).

### 3.2.3 Bismuthene

Bismuth (Bi) is the last and heaviest element of Group-VA [123]. Bulk bismuth is metallic and diamond-shaped in natural state. Studies have shown that when the number of layers is reduced to below 22, bismuth will show the property of topological insulator; when the number of layers is less than 8, it will show the quantum spin Hall phase, and in the case of single layer Bi shows the property of semiconductor [70]. There are many allotropes of Bi such as α-Bi and β-Bi, etc. Like other materials mentioned in this paper, bismuth atoms are also connected with the three atoms around them through covalent bonds, thus forming a low-buckled honeycomb structure as shown in Fig. 6(a)-(b) [36, 124]. The lattice constant is 4.54 Å, and the Bi-Bi bond length is 3.05 Å [62]. The direct bandgap range of bismuth is from 0 to 0.55 eV [60], filling the gap between graphene and black phosphorus. Compared with black phosphorus, Bi is more difficult to be oxidized in the atmosphere, and has a carrier mobility of several thousand, so it is considered as a more promising optoelectronic material [36].

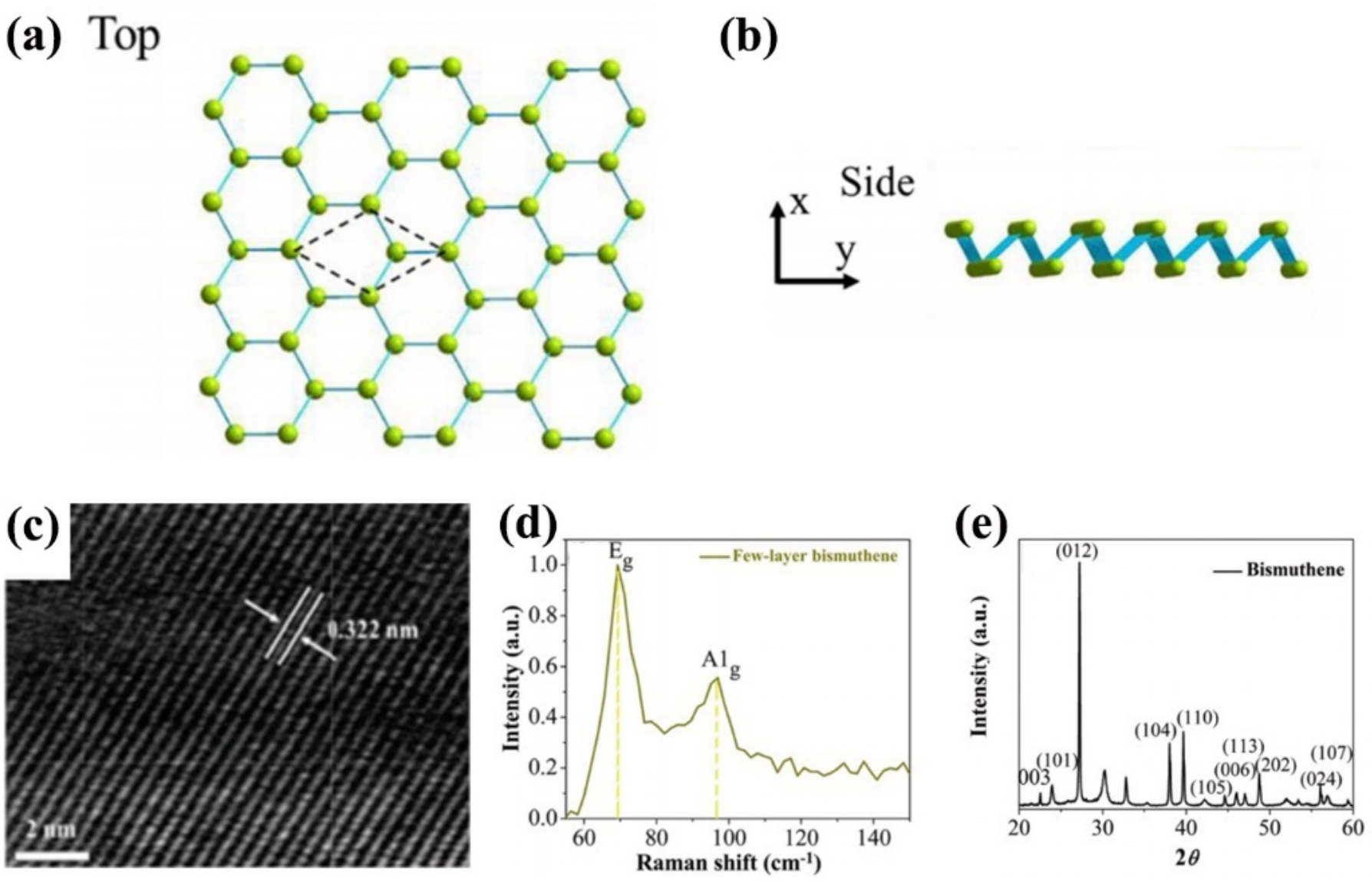

Fig. 6. (a)-(b) Top view and side view of the structure of bismuthine. (c) The HRTEM image of bismuthine. (d) Raman spectroscopy of bismuthine (e) The XRD image of bismuthine. Reproduced with permission from Ref. [125] (a)-(b), Ref. [124] (c), Ref. [123] (d) and Ref. [63] (e).

### 3.3 Dual-elemental analogs

#### 3.3.1 Indium selenide

Indium selenide is a new kind of two-dimensional sulfide semiconductor with high mobility [126]. By means of X-ray diffraction study, it is found that there are about five dielemental compounds in indium selenide melt, among which InSe and $In_2Se_3$ have been deeply studied due to their unique structure [127]. The structure of InSe single crystals is composed of superimposed atomic planes, each of which contains four covalent bond flakes of Se-In-In-Se sequences [128], spaced at 0.83 nm apart. $In_2Se_3$ is composed of vertically superimposed Se-In-Se-In-Se sandwich structure [129, 130]. Studies have shown that there are at least five crystal types in $In_2Se_3$ at different temperatures (i.e. α, β, γ, δ, κ). What more interesting is that InSe is the indirect bandgap (bandgap range from 1.2 eV to 1.4 eV) and $In_2Se_3$ is the direct bandgap. In addition, indium selenide has rhombohedral crystal cells, whose lattice constants are a = b =4.05 Å and c =25.32 Å. As shown in Fig. 7(e), obvious Raman shifts at 115.1, 175.8 and 224.4 $cm^{-1}$ were recorded.

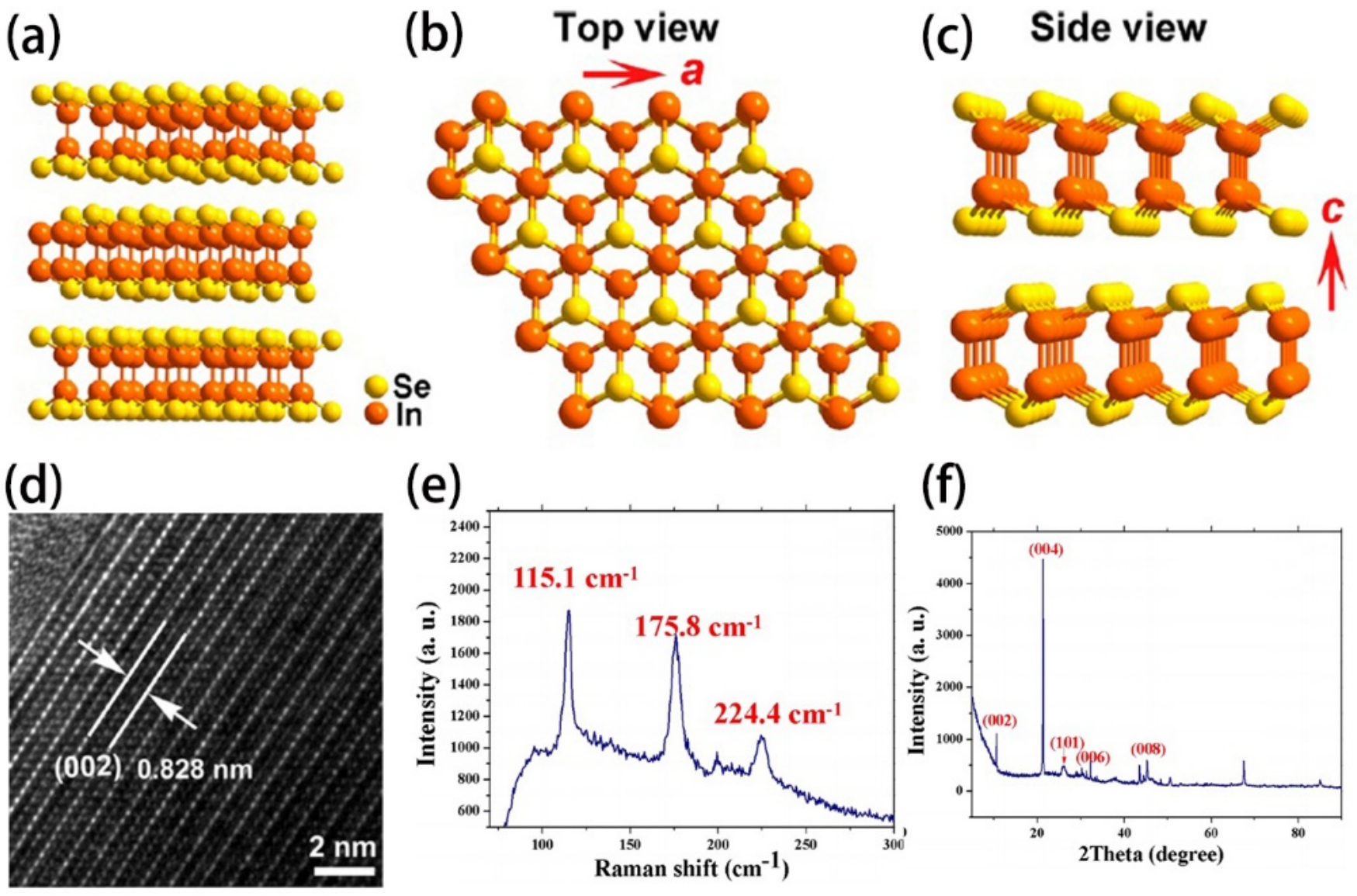

Fig. 7. (a)-(c) The structure of InSe and the top view and side view. (d) The HRTEM image of InSe. (e) Raman spectroscopy of InSe. (f) The XRD image of InSe. Reproduced with permission from Ref. [95] (a)-(d) and Ref. [83] (e)-(f).

### 3.3.2 Tin sulfide

Tin sulfide is a monochalcogenide of Group-VA, which has a rhomboid lattice structure similar to that of black phosphorus [131], and its lattice constant is a =4.33 Å, b=11.19 Å, c=3.98 Å [67]. As shown in Fig. 8(b), the HRTEM image shows the interplanar spacing of atoms of tin sulfide in the plane, which are 0.32 nm (210) and 0.30 nm (111). The most distinctive feature of tin sulfide is that it has both direct and indirect bandgaps, ranging from 1.3 eV to 1.6 eV and 1.0 eV to 1.35 eV, respectively. There is only weak van der Waals force and no suspended bond in SnS, so it has a chemically inert surface. This gives it stable chemical properties. Meanwhile, tin sulfide has a better nonlinear absorption coefficient ($50.5\times10^{-3}$ cm/GW) than black phosphorus. In addition, the non-toxicity of tin sulfide and abundant reserves of its elements are also great advantages of its large-scale application [132].

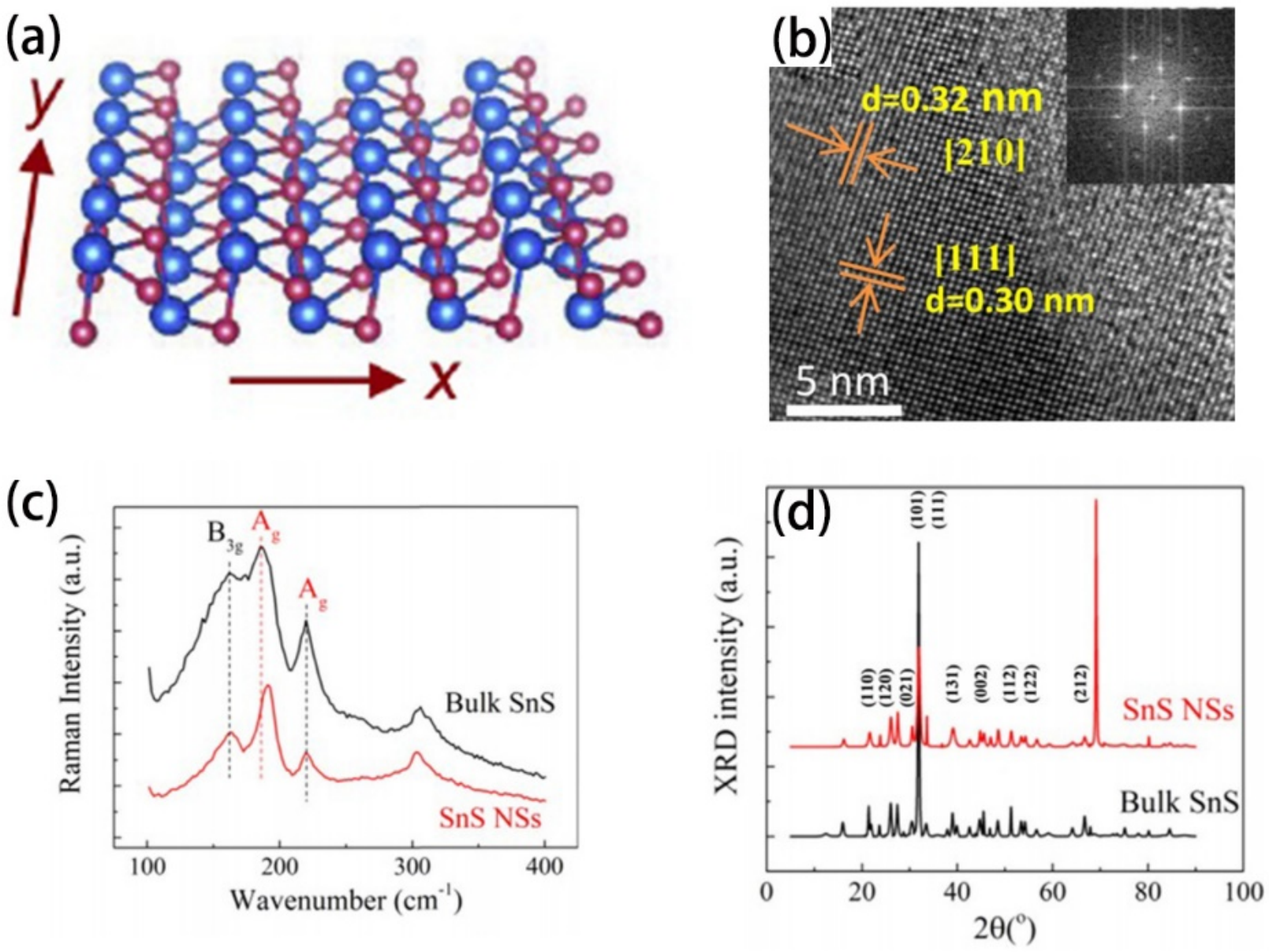

Fig. 8. (a) The structure of SnS. (b) The HRTEM image of SnS. (c) Raman spectroscopy of bulk SnS and SnS nanosheets (SnS NSs). (d) The XRD image of SnS. Reproduced with permission from Ref. [131] (a), and Ref. [133] (b)-(d).

## 4. The performance of black phosphorus and its analogs as saturable absorbers of passively mode-locked fiber lasers

In the previous section, we introduced the properties of black phosphorus and its analogs and found that the analogs are more stable and have a wider adjustable bandgap than black phosphorus, which enables the saturable absorbers made of the analogs to have better stability and wider working ranges. And the advantage of black phosphorus is that it is easier to make. For these reasons, scientists carried out the research simultaneously of black phosphorus and its analogs as saturable absorbers. The scientists focused on the studying of the photoelectric and nonlinear optical properties of black phosphorus and its analogs, such as adjustable bandgap, low saturation strength, fast carrier mobility and small volume, which allow them to be made into saturable absorbers with large modulation depth and flexible operating ranges. It is also worth noting that black phosphorus and its analogs generally have the advantage of being small in size, which makes it easy to be prepared into various forms of saturable absorbers to meet various requirements. Because of their outstanding properties, the passively mode-locked fiber lasers with saturable absorbers made of black phosphorus and its analogs can obtain shorter pulses (up to picosecond and even femtosecond magnitude). Fig. 9 shows a typical structure of the passively mode-locked fiber laser using black phosphorus saturable absorber (BP-SA). In this section, we will talk about the performance of passively mode-locked fiber lasers working at different wavelength and using black phosphorus and its analogs as saturable absorbers.

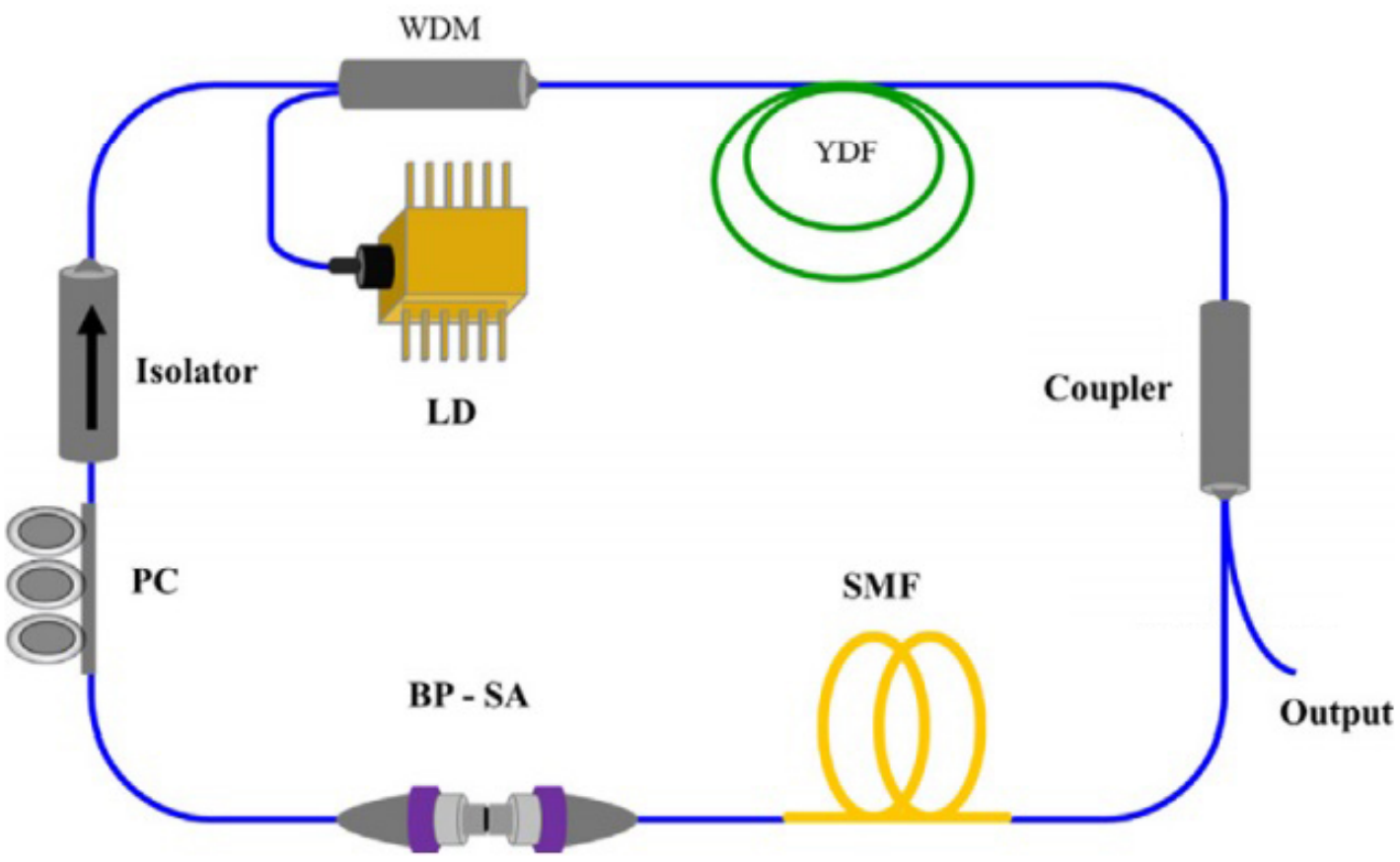

Fig. 9. The schematic of passively mode-locked Er-doped fiber laser using BP-SA. Reproduced with permission from Ref. [39].

### 4.1 Working at ~1 μm

Fiber lasers working near 1.0 μm have been widely studied due to their applications in fiber communication, laser guidance, frequency-doubling laser light source and pumping light source [134]. Table 2 shows the performance of passively mode-locked fiber lasers operating at 1 μm and using BP and its analogs as SA. Li et al. [47] achieved mode-locking operation at high repetition rate of 46.3 MHz at central wavelength of 1030.6 nm. Zhou et al. [135] reported a mode-locked nanosecond fiber laser, which uses BP-SA fabricated by electrochemical delamination exfoliation method, can stably switch the state between of bright and dark pulse. The fiber laser has output power of 8.62 mW in the central wavelength of 1067.1 nm when working at dark pulse state. Harun et al. [136] built a Yb-doped fiber laser using BP-SA with output power up to 80 mW. Ge et al. [123] obtained the ultrashort pulse with pulse width of 76.62 ps as shown in Fig. 10(a) and (e). Tsang et al. [129] explored further applications of α-$In_2Se_3$ in the field of ultrafast photonics and achieved mode-locking operation of 252 fs in Yb-doped fiber lasers. Its central wavelength, average maximum output power and repetition rate are 1060.6 nm, 2 mW and 14.09 MHz respectively.

Table 2 Performance summary of passively mode-locked fiber lasers working 1 μm using BP and its analogs as saturable absorbers.

| Materials | Fabrication Methods | Gain medium | $\alpha_s$ [%] | $I_{sat}$ [MW/cm$^2$] | $\alpha_{ns}$ [%] | λ [nm] | $t_{min}$ | Output Power | RF [MHz] | Ref. |
|---|---|---|---|---|---|---|---|---|---|---|
| BP | LPE | Yb | 8.5 | 2.95 | 1 | 1030.6 | 5 ps | - | 46.3 | [47] |
| BP | ME | Yb | 7.5 | 0.35 | 57.5 | 1033.76 | 3.27 ps | - | 10 | [39] |
| BP | electrochemic--al delamination exfoliation | Yb | - | - | 26.1 | 1063.3 | 386.3 ns | 2.63 | 0.39 | [135] |
| BP | LPE | Yb | 12.5 | 27.9 | - | 1064.6 | 51 ps | - | - | [134] |
| BP | electrochemic--al | Yb | - | - | 26.1 | 1067.1 | 68.4 ns | 8.62 | 0.39 | [135] |
| | delamination exfoliation | | | | | | | | | |
| BP | ME | Yb | 8 | 0.35 | 57 | 1085.58 | 7.54 ps | 80 | 13.5 | [136] |
| Bismuthene | sonochemical exfoliation | Yb | 2.2 | 13 | - | 1034.4 | 30.25 ps | - | 21.74 | [63] |
| Bismuthene | - | Yb | 1 | 2.4 | - | 1035.8 | - | - | 21.74 | [123] |
| α-$In_2Se_3$ | LPE | Yb | - | - | - | 1060.6 | 252 fs | 2 | 14.09 | [129] |
| InSe | LPE | Yb | 4.2 | 15.6 | - | 1068.36 | 1.37 ns | 16.3 | 1.76 | [84] |

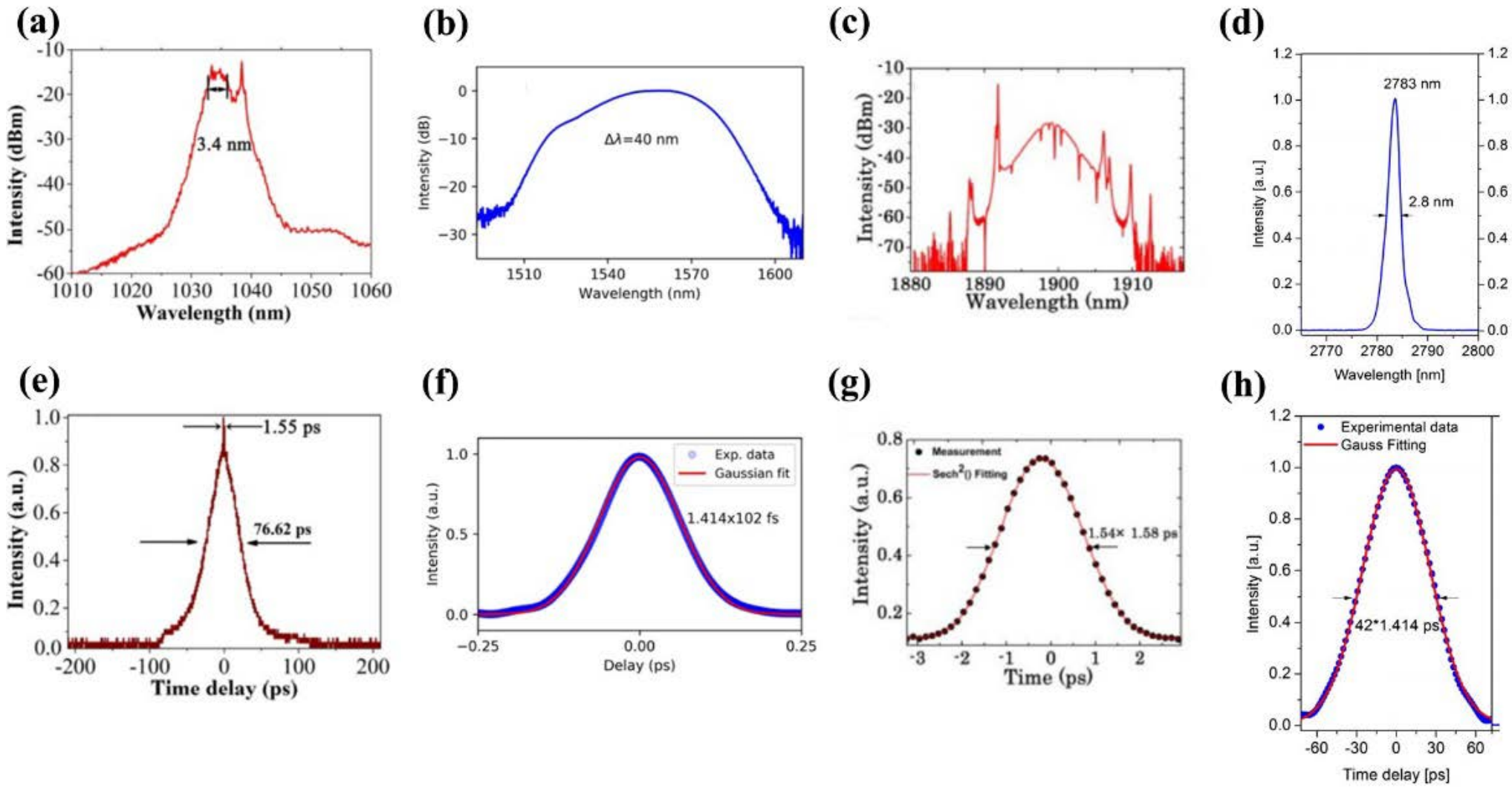

Fig. 10. (a)-(d) Optical spectrum of fiber lasers working at 1, 1.5, 2, 3 μm, respectively. (e)-(h) Autocorrelation trace of fiber lasers working at 1, 1.5, 2, 3 μm, respectively. Reproduced with permission from Ref. [123] (a), Ref. [137] (b), Ref. [138] (c), Ref. [139] (d), Ref. [123] (e), Ref. [137] (f), Ref. [138] (g) and Ref. [139] (h).

### 4.2 Working at ~1.5 μm

In the past decades, many researchers have been studying the output performance of passively mode-locked fiber lasers working near 1.5 μm, which makes the development of fiber lasers near 1.5 μm more mature [140-145]. 1.5 μm is the optical communication window of quartz fiber [146]. However, it was not until 2015 that high-quality few-layered black phosphorus was successfully prepared and its transmittance was found to increase with light intensity. Researchers began to use black phosphorus and its analogs to make saturable absorbers and apply them in passively mode-locked of fiber lasers. Fortunately, due to the fascinating properties of these two-dimensional materials, the research in this field continues to deepen [18, 147-151]. Table 3 shows main achievements to date in this field.

Recently, in addition to black phosphorus, its analogs have also been reported due to their excellent properties of SA [152-159]. Zhu et al. [66] obtained a compact fiber laser with SnS as saturable absorber, and its pulse duration and repetition rate were 1.02 ps and 459 MHz, respectively. Man et al. [130] successfully prepared few-layered $In_2Se_3$ by using physical vapor deposition method and fabricated it as the saturable absorber. By applying it in Er-doped fiber laser, high power output of 121.2 mW is achieved. Hasan et al. [79] reported a long-term stable BP-based mode-locked femtosecond fiber laser, it's pulse duration of 102 fs as shown in Fig. 10(b) and (f). Their research result is the best performance among all reported mode-locked fiber lasers based on BP and its analogs in the 1.5 μm region and showed that black phosphorus has great potential as an excellent candidate for long-term stable ultrashort pulse generation. Zhao et al. [152] and Zhang et al. [151] obtained wavelength tunable ultrafast fiber lasers from 1529 nm to 1592 nm and 1549 nm to 1575 nm, respectively. Their study provides an effective solution for the application of the infrared or mid-infrared ultrafast fiber laser which is widely tunable.

Table 3 Performance summary of passively mode-locked fiber lasers working ~1.5 μm using BP and its analogs as saturable absorbers.

| Materials | Fabrication Methods | Gain medium | $\alpha_s$ [%] | $I_{sat}$ [MW/cm$^2$] | $\alpha_{ns}$ [%] | λ [nm] | $t_{min}$ | Output Power | RF[MHz] | Ref. |
|---|---|---|---|---|---|---|---|---|---|---|
| BP | LPE | Er | 10.03 | 14.98 | 9.97 | 1555 | 102 fs | - | 23.9 | [137] |
| BP | water-exfoliated method | Er | 1 | 221 | - | 1557.8 | 1.2 ps | - | 6.317 | [140] |
| BP | - | Er | 3.31 | 12.5 | 73.6 | 1558.14 | 2.18 ps | - | 15.59 | [141] |
| BP | - | Er | - | - | - | 1558.7 | 786 fs | - | 14.7 | [142] |
| BP | LPE | Er | 0.35 | - | - | 1558 | 700 fs | 1.5 | 20.82149 | [143] |
| BP | LPE | Er | 21.00 | 12 | 40 | 1559.5 | 670 fs | - | ~8.77 | [144] |
| BP | ME | Er | 7.00 | 0.25 | 58 | 1560.7 | 570 fs | 5.1 | 6.88 | [145] |
| BP | ME | Er | 7.75 | - | 10 | 1560 | 580 fs | - | 15.2 | [147] |
| BP | LPE | Er | 0.8 | - | - | 1561 | 1.438 ps | - | 5.268 | [18] |
| BP | ME | Er | 6.90 | 0.25 | - | 1561 | 2.66 ps | - | 1 | [148] |
| BP | LPE | Er | ~0.3 | - | - | 1562 | 1.236 ps | - | 5.426 | [18] |
| BP | - | Er | - | - | - | 1564.6 | 690 fs | - | 3.47 | [149] |
| BP | LPE | Er | 10.90 | 25 | - | 1566.5 | 940 fs | 5.6 | 4.96 | [150] |
| BP | ME | Er | 8.10 | 6.55 | - | 1571.45 | 946 | - | 5.96 | [53] |
| | | | | | | | fs | | | |
| BP | LPE | Er | 10 | 8.3 | - | 1576.1 | 403.7 fs | 1.9 | 34.27 | [134] |
| BP | LPE | Er | 10.1 | 9.27 | - | tunable | 280 fs | - | - | [151] |
| BP | LPE | Er | 0.76 | 5.1 | 22.64 % | tunable | 0.9 ps | - | 5.65 | [152] |
| BPQDs (black phosphorus quantum dots) | LPE | Er | 8.10 | - | - | 1561.7 | 0.88 ps | - | 5.47 | [153] |
| BPQDs | LPE | Er | 9 | 1.5 | - | 1562.8 | 291 fs | - | 10.36 | [154] |
| BPQDs | solvothermal synthesis | Er | 36 | 3.3 GW $cm^{-2}$ | - | 1567.5 | 1.08 ps | - | 15.25 | [155] |
| Tellurene | LPE | Er | 11.86 | 44.65 GW $cm^{-2}$ | - | 1556.57 | 879 fs | 3.45 | 15.45 | [115] |
| Tellurene | LPE | Er | 10.5 | 26 GW $cm^{-2}$ | - | 1556.57 | 879 fs | 3.45 | 15.45 | [115] |
| Tellurene | LPE | Er | 27 | 78.14 GW $cm^{-2}$ | - | 1556.57 | 879 fs | 3.45 | 15.45 | [115] |
| Tellurene | LPE | Er | 35.64 | 1.06 | 25.91 | 1558.8 | 1.03 ps | 3.69 | 3.327 | [156] |
| Tellurene | LPE | - | 5.06 | 34.3 | - | 1563.97 | 38.5 ps | 106.6 | 12.17 | [118] |
| Tellurene | LPE | Er | 0.97 | - | 42.30 | 1565.58 | 21.45 ps | 17.44 | 5.0378 | [82] |
| Tellurene | LPE | - | 5.06 | 34.3 | - | 1573.97 | 5.87 ps | 23.61 | 12.17 | [118] |
| Antimonene | vapor deposition | Er | 11.77 | 9.17 | - | 1562.64 | 753 fs | - | - | [157] |
| Antimonene | MSD | Er | 25 | 28 | - | 1557 | 544 fs | 21.76 | 17.15 | [48] |
| Antimonene | LPE | - | 3.96 | 14.25 GW $cm^{-2}$ | - | 1557.68 | 552 fs | - | 10.27 | [80] |
| Antimonene | LPE | - | 19.72 | 15.10 GW $cm^{-2}$ | - | 1557.68 | 552 fs | - | 10.27 | [80] |
| Bismuthene | LPE | Er | 7.7 | 16 | - | 1530.4 | 1.75 ps | - | 4 | [158] |
| SnS | sol-hydrothermal method | Er | 5.4 | 66.3 | - | 1530.6 | 1.29 ps | - | 5.47 | [65] |
| Bismuthene | LPE | Er | 2.5 | 113 | 66.3 | 1531 | 1.303 | - | 4 | [36] |
| | | | | | | | ps | | | |
| Bismuthene | - | Er | 2.5 | 110 | - | 1555.7 | - | - | 4 | [159] |
| Bismuthene | sonochemical exfoliation method | Er | 2 | 0.3 | - | 1557.5 | 621.5 fs | - | 22.74 | [62] |
| Bismuthene | sonochemical exfoliation method | Er | 2.03 | 30 | - | 1559.18 | 652 fs | - | 8.83 | [61] |
| Bismuthene | sonochemical exfoliation method | Er | 5.60 | 48.2 | 62.3 | 1561 | 193 fs | 5.6 | 8.85 | [60] |
| α-$In_2Se_3$ | ME | Tm | 14.6 | 0.4 KW/$cm^2$ | - | 1503.8 | 5.79 ps | 1.242 | 6.98 | [76] |
| α-$In_2Se_3$ | LPE | Er | - | - | - | 1550 | 215 fs | 2 | 7.31 | [129] |
| $In_2Se_3$ | PVD | Er | 18.75 | 6.8 | 18.89 | 1559.4 | 14.4 ns | 122.4 | 1.71 | [130] |
| α-$In_2Se_3$ | LPE | Er | - | - | - | 1560.08 | 215 fs | 2 | 7.31 | [129] |
| α-$In_2Se_3$ | MSD | Er | 4.5 | 7.3 | 21.9 | 1565 | 276 fs | 83.2 | 40.9 | [160] |
| SnS | LPE | - | 36.40 | ~34.8 GW $cm^{-2}$ | 27.7 | 1560 | 656 fs | - | 8.37 | [133] |
| SnS | hydrothermal method | Er | 5.8 | 52.2 | ~35 | 1562.4 | 1.02 ps | - | 459 | [66] |

Table 4 Performance summary of passively mode-locked fiber lasers working 2 μm and 3 μm using BP and its analogs as saturable absorbers.

| Materials | Fabrication Methods | Gain medium | $\alpha_s$ [%] | $I_{sat}$ [MW/$cm^2$] | $\alpha_{ns}$ [%] | λ [nm] | $t_{min}$ | Output Power | RF [MHz] | Ref. |
|---|---|---|---|---|---|---|---|---|---|---|
| BP | ME | Tm | 9.8 | - | 78.1 | 1898 | 1.58 ps | - | 19.2 | [138] |
| BP | ME | Tm | 4.1 | - | 52.2 | 1910 | 739 fs | 1.5 | 36.8 | [161] |
| Chem-Te | LPE | Tm | 13.71 | 80.7 | 32.09 | 1971 | 890 fs | 5.35 | 11.17 | [162] |
| Antimonene | MSD | Tm | 48 | 32 | - | 1892 | 972 fs | 36.1 | 15.5 | [48] |
| α-$In_2Se_3$ | MSD | Tm | 6.9 | 10.6 | 28.8 | 1932 | 1020 fs | 112.4 | 15.8 | [160] |
| BP | ME | Er | 19 | - | - | 2783 | 42 ps | 613 | 24 | [139] |
| BP | LPE | $Ho^{3+}/Pr^{3+}$ | 41.20 | 3.767 | 7.6 | 2866.7 | 8.6 ps | 87.8 | 13.987 | [163] |
| BP | - | Er | 7.70 | - | 14 | 3489 | 213 fs | 40 | 28.91 | [164] |

### 4.3 Working at ~2 μm and ~3 μm

The fiber lasers operating near 2 μm most use thulium as the gain medium. Meanwhile, it is safe for human eyes, the 2 μm laser thus widely used in meteorological monitoring, laser ranging, lidar, remote sensing and other fields [165]. In addition, water molecules near 2 μm has a strong mid-infrared absorption peak, the surgery using this band laser is conducive to speed up blood coagulation and reduce surgical trauma. Mid-infrared fiber laser in the medical and life sciences has an important application. Abramski et al. [161] first reported a thulium-doped fiber laser using BP-SA and obtained 739 fs mode-locked pulse. Yang et al. [160] used MSD method to make α-$In_2Se_3$ into a saturable absorber with broadband saturation absorption characteristics, and they applied it to a passively mode-locked fiber laser to obtain a mode-locked pulse with pulse width of 1020 fs and output power of 112.4 mW. In addition, Nie et al. [162] used liquid phase peeling technique to produce a novel chemically synthesized tellurium thin film (Chem-Te) and first reported its saturable absorption characteristics in a 2 μm fiber laser. Fig. 10(c) and (g) show the optical spectrum and autocorrelation trace of fiber laser working at ~2 μm measured by Jiang et al.

Fiber lasers working at ~3 μm have potential applications in biological, medical and other fields. Meanwhile, these lasers can also be used as pumping light source for medium and far infrared laser. Fluoride glass is generally as the fiber substrate of the passively mode-locked fiber lasers working at ~3 μm because the laser emission near 3 μm requires the substrate material to have low phonon energy and high optical head pass rate. Shen et al. [164] first applied saturable absorber made of black phosphorus in an erbium-doped fiber laser in the region of 3 μm, and obtained a pulse width of 213 fs. At the same time, the output power of the laser could reach 40 mW. Qian et al. [139] used ME method to prepare multilayer black phosphorus and applied it to ultrafast fiber laser, obtaining the maximum average output power of up to 613 mW as shown in Fig. 10(d) and (h). Liu et al. [163] used the co-doped $Ho^{3+}/Pr^{3+}$ fluoride fiber as the gain medium and black phosphorus as the saturable absorber to obtain the passively mode-locked output, which has pulse width of 8.6 ps, output power of 87.8 mW and repetition rate of 13.987 MHz. Table 4 shows data we collected about the performance of the fiber lasers working at 2 and 3μm.

## 5. Applications of black phosphorus and its analogs

At present, black phosphorus and its analogs are widely used in many fields [102], such as photodetector, optical modulators, solar cells, transistors [208], electrochemical energy storage devices, piezoelectric devices [117], as well as in drug delivery, bioimaging, photothermal therapy, photodynamic therapy, photocatalyst and sensors, as shown in Fig. 11. Of course, this is due to their excellent material properties. For example, BP is suitable for photodetector [27] because it has a strong impact to the photon from near ultraviolet to infrared wavelengths. SnS has great application potential in the solar cells and other optical devices because its bandgap is close to the best forbidden band width of solar cells.

## 5.1 Optoelectronic characteristics

We believe that the application of black phosphorus and its analogs in the field of optoelectronics is worthy of the attention from scientific community [166]. And the photodetector is very representative in this field [167]. The photodetector is a device that converts optical signals into electrical signals, which has a wide range of applications in daily life, including environmental monitoring, medical imaging, optical communication, security and military applications [168]. Two-dimensional materials such as black phosphorus generally have strong photo-material interactions, making them promising candidates for high performance photodetector applications. Ozyilmaz et al. [169] proved that black phosphorus is an excellent material for ultraviolet photodetectors with specific detectivity up to $3 \times 10^{13}$ Jones. Photodetectors made of black phosphorus and its analogs have excellent performance, including ultrabroad detecting band, ultra-high light absorption efficiency and fast photo-response. Studies have shown that black phosphorus and its analogs have a wide detection band from near ultraviolet to long-wave infrared [170]. Two-dimensional layered materials such as black phosphorus and its analogs interact strongly with photon. The quantum constraint of vertical direction leads to a spike in the state density near the edge of the conduction and valence band, which increases the possibility of excited free electron-hole pairs when the energy of the incident photon approaches the band gap, resulting in higher photon absorption [171]. Cai et al. explored the difference in photon absorption between the two crystal axes caused by the anisotropy of black phosphorus and its analogs, and for BP of 70 nm, the light absorption varies between 40% and 10% from the armchair direction to the zigzag direction [172]. BP showed a wide and rapid response from the visible region to the near-infrared region [173]. Buscema et al. [174] found that the photodetector made of few-layered black phosphorus showed a response to the wavelength from the visible region to 940 nm and the rise time of about 1 ms, which proved its broadband and rapid detection, and its responsivity reached 4.8 mA/W. Due to the weak interlayer forces of two-dimensional materials, the lattice mismatch requirements between different two-dimensional materials are lower than those of conventional semiconductor heterojunctions. Bao et al. [175] reported a graphene-BP heterostructure photodetector, which shows long-term stability and ultra-high photon response ($3.3\times10^3$ $AW^{-1}$) at near-infrared wavelength (1550 nm). Zhang et al. [176] fabricated the $Te_x@Se_y$ with roll-to-roll nanotubes heterojunctions by epitaxial growth of Se on the Te nanotubes and firstly applied it to the photodetectors. Javey et al. [177] found people can increase absorption in semiconductors using $Al_2O_3/Au$ optical cavity substrate and make photodetectors using Te by this way. They measured the peak responsiveness of Te photoconductor when the thickness of $Al_2O_3$ was 550 nm, 150 nm and 350 nm respectively, and this parameter changed from 13 to 8 A/W. Therefore, it was concluded that the performance of photodetectors made by Te could be changed through changing the thickness of the $Al_2O_3/Au$ optical cavity substrate.

In addition to photodetectors, optical modulators are also important optoelectronic devices [178]. Black phosphorus is easy to manufacture high performance all-optical modulators due to their excellent nonlinear optical properties. At present, the BP-based optical modulator mainly uses the Stark effect and the Pauli blocking effect, the former can effectively reduce the bandgap under the vertical electric field, while the latter can transfer the absorption edge to a higher energy [179]. Zhang et al. [180] investigated an all-optical modulator using phosphoene that utilizes a Mach-Zehnder interferometer and achieves stability under environmental conditions. Helmy et al. [181] demonstrated that a BP-based optical modulator can achieve better maximum absorbability and power efficiency than graphene.

In addition, as solar energy is the new clean and environmentally friendly energy, people try to use two-dimensional materials to make solar cells, so as to make better use of solar energy [182, 183]. Solar cells are devices that convert light energy directly into electricity through the optoelectronic or photochemical effect. Syed Mubeen et al. [184] believed that SnS thin films could be used as an effective photocathode with the maximum photocurrent density up to 12 mA $cm^{-2}$. Shapter et al. [185] and Blom et al. [186] summarized the current research status on solar cells of black phosphorus. Yu et al. [187] carried out a research on improving the performance of solar cells. They combined the photocathode based on BPQDs (black phosphorus quantum dots) into DSSCs (Dye sensitized solar cells), which increased the electron density of DSSCs and improved the electron transport performance of DSSCs. Compared with photoanode of $TiO_2$ films, Chu et al. [188] enhanced optoelectronic conversion of 38% using BP/$TiO_2$ composites.

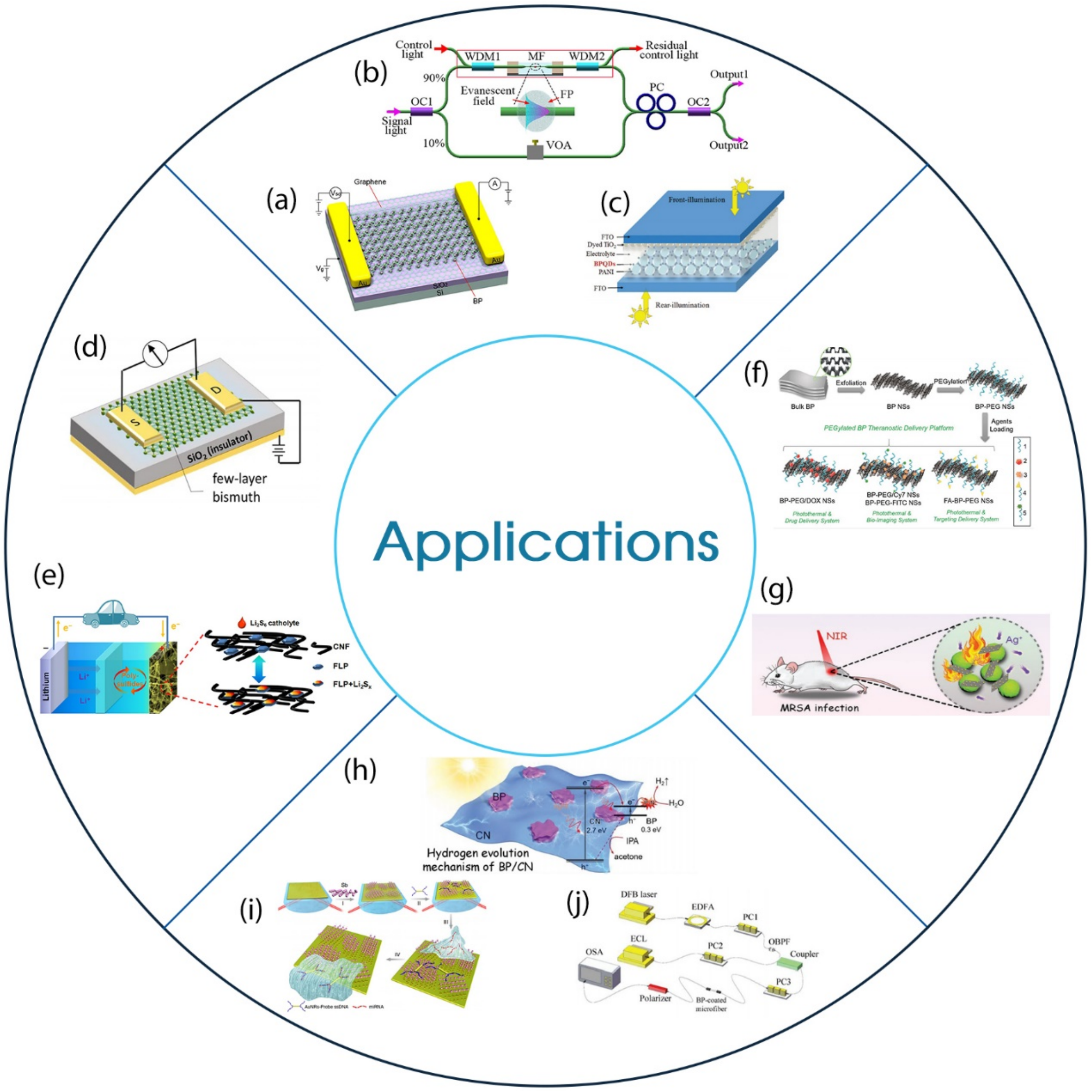

Fig. 11. (a) Three-dimensional schematic diagram of the graphene-BP heterostructure photodetector. (b) Experimental setup of the Mach-Zehnder interferometer-based all-optical modulator. (c) The schematic of the bifacial n-type dye-sensitized solar cell with BPQDs (black phosphorus quantum dots). (d) The schematic of FETs based on Bi film. (e) Schematic of the few-layer phosphorene and carbon nanofiber matrix used as the host for the lithium polysulphide catholyte. (f) Schematic representation of the PEGylated BP theranostic delivery platform. 1: PEG-$NH_2$ (surface modification), 2: DOX (therapeutic agents), 3: Cy7-$NH_2$ (NIR imaging agents), 4: folic acid-PEG-$NH_2$ (targeting agents), 5: fluoresceine isothiocyanate-PEG-$NH_2$ (fluorescent imaging agents). (g) the schematic of synergistic antibacterial effect. (h) Schematic of the charge transfer process and mechanism of photocatalytic $H_2$ emission. (i) The process of a Sb-based miRNA sensor fabrication and detect miRNA use it. (j) The schematic of Kerr switch. Reproduced with permission from Ref. [175] (a), Ref. [180] (b), Ref. [105] (c), Ref. [125] (d), Ref. [189-193] (e)-(i) and Ref. [105] (j).

**5.2 Electronics**

In the field of modern electronics, the transistor, as an indispensable basic unit of integrated circuit, has been widely studied. BP and its analogs are considered excellent materials for making transistors [194, 195]. Because they all have a limited bandgap, which allows the transistors they make to have a high pass ratio. Wu et al. [22] reported a field-effect transistor (FETs) made by Te, which had good air stability, an on/off ratio of up to $10^6$, and a field effect mobility of about 700 $cm^2 V^{-1} s^{-1}$, and shown another transistor with the conduction density of 1 A $mm^{-1}$. Lu et al. [194] studied the performance limits of single-layer BP Schottky barrier transistors below 10 nm using *ab initio* quantum transport simulations for the first time. Hao et al. [125] explored the feature of p-type field-effect transistors by studying the Bi-film FETs, which have the large carrier mobility of 220 $cm^2 V^{-1} s^{-1}$. Two-dimensional materials such as BP and its analogs have high elasticity and flexibility, which makes them suitable for flexible electronic products. Based on this development prospect, we can foresee that the flexible research of transistors made from these materials will have a broad prospect.

Electrochemical energy storage devices (EESDs) are the device for converting chemical energy into electrical energy. Because their structure is simple, easy to carry, easy to charge and discharge operation, not affected by the outside climate and temperature, stable and reliable performance, in the modern social life in all aspects play a great role [196-198]. The flexibility, large surface area and good electrical conductivity of BP and its analogs make them a promising electrode material of EESDs [72]. Koratkar et al. [189] attempted to insert the phosphorus nanosheets into the cathode matrix of lithium sulfur battery to significantly improve the cycle life of lithium sulfur battery and improve the utilization rate of sulfur in the battery. Xiu et al. [199] summarized and analyzed the latest research results of black scale and other materials as the positive pole of lithium-based lithium batteries. Guo et al. [200] enhanced the properties of sodium ion batteries, such as structural stability and cyclic stability, by conducting structural phase transition in SnS. Yun et al. [201] applied SnS nanoparticles to the anode of the $Na^+$ battery, and found that the specific energy, power value and first Coulombic efficiency were ~256 Wh $kg^{-1}$,471 W $kg^{-1}$, 90%, respectively.

**5.3 Biomedicine**

In the field of biomedicine, 2D materials have potential applications in photothermal therapy, photodynamic therapy, bioimaging, drug delivery and so on [202]. For example, BP can produce high photothermal conversion efficiency and thus can be used as a photosensitizer in photothermal therapy and photodynamic therapy [203]. By processing 2D materials to the nanometer level and making nanometer probes, they can be used for multi-peak imaging of tumors [204]. BP have certain antibacterial properties and can be used as a clinical platform for against bacteria [191]. Due to the limited space, we mainly present the application status of BP and its analogs in drug delivery.

2D materials have the properties of high drug loading because of their large surface area. Furthermore, 2D BP has low cytotoxicity [205] and degradability, making it an ideal drug delivery platform [21]. A good drug delivery system enables drugs to be released more intensively and efficiently at the target location, significantly reducing the side effects of drugs and improving the efficiency of drug [203]. Kim et al. [206] assembled Au and $\gamma$-$Fe_2O_3$ nanoparticles on BP nanosheets (BP-NSs) to develop a biocompatible composite material, the property that enables this material to be used for drug delivery. Zhang et al. [207] summarized the latest advances in medical diagnosis and treatment of few-layered BP. Mei et al. [190] prepared BP nanosheets by ME method, then used polyethylene glycol–amine (PEG-$NH_2$) to improve their biocompatibility and physiological stability, and proved that the drug delivery platform they developed could effectively load doxorubicin (DOX) and cyanine7 (Cy7). Their study shows the medical properties of PEGylated BP-NSs for the first time and the feasibility of using BP nanosheets for medical treatment.

**5.4 Other fields**

BP and its analogs are not only widely used in the fields mentioned above, but considered ideal and attractive materials in many fields in fact. For example, BP are also used as optical Kerr converters and wavelength converters based on four-wave mixing [105] in the field of optics. We certainly can't go into details here, but we want to present as many of the hottest aspects of current research as we can.

We think what should be paid attention to is the application of black phosphorus as the photocatalyst. Based on the light collected, black phosphorus produces electron-hole pairs that can redox some substance such as water. Majima et al. [208] reported that black phosphorus flakes promoted the release of $H_2$ in water decomposition reactions, with apparent quantum efficiencies of 8.7% and 1.5% under visible (420 ± 5 nm) and near-infrared light (780 ± 5 nm), respectively. Yu et al. [192] prepared a kind of nanocomplexes consisting of black phosphorus and graphitic carbon nitride (CN). They placed BP/CN in water containing IPA to produce hydrogen at a release rate of 786 μmol $h^{-1}$ $g^{-1}$. At the same time, they further explored the photocatalytic principle of BP/CN. Wang et al. [209] summarized the latest progress of BP/CN in photocatalysis. Compared with black phosphorus, this hybrid has better light collection ability and photocatalytic performance. Yang et al. [210] summarized some common performance improvement methods for two-dimensional material catalysts such as black phosphorus.

Black phosphorus and its analogs also show unique advantages in sensor applications. Due to their large surface volume ratio and environmental sensitivity, they have been widely used in gas sensors, biosensors and humidity sensors [211]. Rong et al. [211] summarized the current application status and future prospects of black phosphorus in the field of sensors. Bao et al. [142] explored the application of Sb-based sensors in MicroRNA detection. Its detection limit is 2.3-10,000 times higher than existing sensors, providing a better choice for cancer diagnosis and detection. There is no doubt that these materials' high sensitivity, repeatability and stability pave the way for industrial applications, safety equipment and health sensing.

## 6. Conclusion and outlook

As we introduced above, black phosphorus and its analogs have excellent properties, and some achievements have been made on their research [23, 206, 212-214]. But from our perspective, the following issues still need to be tackled in the future:

(1) In terms of preparation method, the existing methods have various problems, for example, the yield of monolayer material prepared by ME is low. LPE method will introduce internal or external defects into 2D materials, which will adversely effect on the application in the next step. And researchers have not yet found the most suitable substrate material of CVD. Meanwhile, although each method claims to be able to produce a controlled thickness of black phosphorus, researchers cannot obtain black phosphorus at the designated thickness. Therefore, how to improve the existing preparation methods or find other more suitable preparation methods will be researchers' future works.

(2) As described above, black phosphorus is easily oxidized and environmentally unstable. If it is exposed to air, its surface will be corroded layer by layer. Although the degradability of black phosphorus is an advantage in medical applications, in other fields, especially in optics, when researchers use BP as saturable absorbers to generate passively mode-locked pulses, they want the black phosphorus to be more stable so that they can get the steady pulse. Therefore, the preparation of black phosphorus with high stability is also a major research direction [215]. At present, there are two main ways to overcome the degradability of BP, one is surface modification, the other is surface coating [215, 216]. However, the research on the degradation mechanism of black phosphorus caused by environment is still in its infancy. Therefore, the research on the anti-degraded BP is only in the initial stage. In the future, researchers need to go further.

(3) In the material family of black phosphorus and its analogs, researchers find that the properties of black phosphorus have been extensively studied and revealed, but the study of analogs is still in its early stages, especially many questions remain to be solved in the field of application. Because scientists have studied the properties of 2D black phosphorus deeply, the research on its application is far ahead of its analogs. In the future work, researchers should continue to explore the properties, preparation methods and applications of black phosphorus analogs.

(4) Heterogeneous structures based on two-dimensional layered materials usually have better performance than single two-dimensional layered materials [22, 168, 217-221]. At the same time, the heterostructure can also be an ideal saturable absorber for pulsed lasers [222]. Therefore, researchers can try to study heterogeneous integration of black phosphorus and its analogs in order to obtain precisely designed van der Waals heterostructures and thus build more powerful devices such as high-power lasers, high-electron-mobility transistors and bipolar transistors [101].

In this review, we have reviewed preparation methods, material properties, current applications of black phosphorus and its analogs, and the performance of them as saturable absorbers in passively mode-locked fiber lasers. We focus on showing their properties as saturable absorbers in lasers working at different central wavelength, which are demonstrated by the pulse generated by the lasers. As we have seen, passively mode-locked pulse outputs from 1 μm to 3 μm can be achieved using black phosphorus and its analogs as SA. At the same time, we can also notice that BP and its analogs, especially BP, are widely used in industry, medical treatment and other fields. With the development of materials science, we firmly believe that the properties of 2D BP and its analogs will be further discovered, they will be ideal SA and will be wider application in various fields.

**Acknowledgements** We acknowledge the financial support by the China Postdoctoral Science Foundation (No. 2019M651203), Science and Technology Project of the 13th Five-Year Plan of Jilin Provincial Department of Education (Nos.

JJKH20190157KJ and JJKH20190169KJ), Jilin Province Science and Technology Development Plan Project (No. 20190201128JC), and Changchun Science and technology plan project (No. 18SS008).